\documentclass[useAMS,usenatbib]{mn2e}
\usepackage{epsfig}
\usepackage{psfig}
\usepackage{latexsym}
\usepackage{amsmath}
\usepackage{amssymb}
\usepackage{graphics}
\usepackage{verbatim}
\usepackage[usenames]{color}
\usepackage{shortbold}

\newcommand{\INPUT}[1]{}
\newcommand{\cut}[1]{}

\title[Cleaning sky survey databases with Hough transforms \& renewal
strings] {Cleaning Sky Survey Databases using Hough Transform and Renewal
String Approaches}

\author[A.J.\ Storkey, N.C.\ Hambly, C.K.I.\ Williams \& R.G.\ Mann]
{A. J. Storkey$^{1,2}$, N. C. Hambly$^2$, C. K. I. Williams$^1$, R. G. 
Mann$^{2,3}$\\
$^1$School of Informatics, University of Edinburgh,
Forrest Hill, Ediburgh EH1~2QL.\\
$^2$Institute for Astronomy, University of Edinburgh, Blackford Hill,
Edinburgh, EH9~3HJ\\ $^3$National E-Science Centre, South College Street,
Edinburgh, EH8~9AA.}
\date{Accepted 2003 September 12.
      Received 2003 September 10;
      in original form 2003 July 10}

\begin{document}
\maketitle
\begin{abstract}
Large astronomical databases obtained from sky surveys such as the SuperCOSMOS
Sky Survey (SSS) invariably suffer from spurious records coming from
artefactual effects of the telescope, satellites and junk objects in orbit
around earth and physical defects on the photographic plate or CCD. Though
relatively small in number these spurious records present a significant
problem in many situations where they can become a large proportion of the
records potentially of interest to a given astronomer. Accurate and robust
techniques are needed for locating and flagging such spurious objects, and we
are undertaking a programme investigating the use of machine learning
techniques in this context. In this paper we focus on the four most common
causes of unwanted records in the SSS: satellite or aeroplane tracks,
scratches, fibres and other linear phenomena introduced to the plate, circular
halos around bright stars due to internal reflections within the telescope and
diffraction spikes near to bright stars. Appropriate techniques are developed
for the detection of each of these. The methods are applied to the SSS data to
develop a dataset of spurious object detections, along with confidence
measures, which can allow these unwanted data to be removed from
consideration. These methods are general and can be adapted to other
astronomical survey data.
\end{abstract}
\begin{keywords}
astronomical databases: miscellaneous -- catalogues -- surveys -- methods:
data analysis -- statistical
\end{keywords}

\section{Introduction}
Sky surveys in astronomy are a fundamental research resource
\citep{banday_mining_sky}. Surveys form the basis of statistical studies of
stars and galaxies, enabling work ranging in scale from the solar
neighbourhood to a significant fraction of the observable universe. Surveys
are carried out in all wavelength ranges, from high energy gamma rays
\citep{paciesas_batse_catalogue} to the longest wavelength radio atlases
\citep{bock_wide_field_radio_southern_1}. Despite this diversity, there are
certain features common to most digital surveys: pixel images at a given
spatial and spectral resolution are processed using a pixel analysis engine to
generate lists of object detections containing parameters describing each
detection. In most cases, the object detection algorithm has to be capable of
finding a heterogeneous family of objects, for example point--like sources
(stars, quasars); resolved sources (e.g.~galaxies) and diffuse, low
surface--brightness, extended objects (e.g.~nebulae). Object parameters
describing each detection typically include positions, intensities and shapes.
The volume of pixel data required to be processed necessitates totally
automated pixel processing, and of course no imaging system is perfect. These
facts (imperfect image recording and automated, generalised pixel processing)
lead to the problem of spurious object catalogue records in all sky survey
databases, although the exact nature of the spurious objects varies. For
example, direct digital sky surveys suffer less from satellite tracks (though
they do exist) because of the short exposure time needed for charge coupled
device (CCD) arrays compared with photographic plates. In fact because the
satellite tracks tend to be significantly shorter, they are harder to detect
using standard approaches making the developments in this paper more rather
than less important. Infra-red surveys are likely to suffer less from
satellite track problems as the tracks will be a couple of orders of magnitude
fainter in the near-infra-red than in the optical. However it is still likely
that some will be detectable, even if in smaller numbers. Optical artefacts
from telescope design occur independent of the digitisation method.

This paper looks at a class of problems which are the most significant sources
of unwanted records in the SuperCOSMOS Sky Survey (SSS) data. The SSS is
described in a series of papers (\citealt{hambly_supercosmos_1_short} and
references therein). Briefly, the SSS consists of Schmidt photographic plates
scanned using the fast, high precision microdensitometer SuperCOSMOS
(e.g.~\citealt{hambly_supercosmos}). The survey is made from $894$ overlapping
fields in each of three colours (blue, red and near--infrared denoted by the
labels J, R~and~I respectively); one colour (R) is available at two epochs to
provide additional temporal information. Each image contains approximately
$10^9$ 2-byte pixels. The pixel data from each photograph in each colour and
in each field are processed into a file of object detections; each object
record contains parameters describing that object. The SuperCOSMOS pixel
analyser is described in \citet{hambly_supercosmos_2} and references therein.
Also described are some classification and quality flags that are derived for
each object detection, and the deblending algorithm which attempts to
unscramble groups of close or merged objects. Presently, the entire southern
hemisphere is covered, primarily using plates from the UK Schmidt Telescope.
Data and many more details are available online at
\verb+http://www-wfau.roe.ac.uk/sss+.

SSS data take the form of pixel images and object catalogues derived from
them. The SSS database is, like any other astronomical database, subject to
the limitations of its imaging system and pixel processing engine. The SSS
object catalogues are therefore contaminated by spurious object records.
The types of spurious objects fall into three broad classes:
\begin{itemize}
\item Linear features: Satellite tracks, aeroplane tracks, fibres left on the
plates during digitisation and scratches on the emulsion all produce linear
features with varying curvature characteristics and lengths. Scratches and
fibres tend to be short. Satellite tracks can be short or long for a variety
of reasons. Aeroplane tracks usually traverse the plate but often consist of
dashed sections corresponding to a flashing light. Spinning satellites can
also caused dashed tracks.
\item Near-circular elliptical optical artefacts around bright stars due to the internal
reflections within the telescope.
\item Diffraction spikes: linear features which are located (almost)
horizontally and (almost) vertically on the plate in the region of bright
stars.
\end{itemize}
How these features translate into objects in the sky survey catalogue depends
also on the approach of the program which processes the digital picture into
object catalogue records. For the SuperCOSMOS Sky Survey many of the largest
linear features or parts of linear features are
clearly non--astronomical in origin, cannot be processed by the pixel
analyser, and therefore do not give rise to spurious object catalogue records.
The rest tend to be represented in the catalogue as a number of objects lying
along a line. Hence even if a track traversed the whole plate in the original
image, in the derived catalogue data it might only translate into a set of
objects traversing a short section of the original track.

The focus of this paper is on locating objects in an astronomical dataset
derived from or affected by the characteristics listed above, and
distinguishing them from true astronomical objects. Because much work has
usually already been done deriving the object data from images, because in
many cases original image data may not be available, and because of the huge
size of the images involved, we are not considering working with the images
directly, only with the derived datasets.

The paper continues with Section~\ref{sect:spuriousdetail}, giving further
detail of the different sorts of spurious data which might be encountered.
Some standard image processing techniques from the computer vision community
are introduced in Section~\ref{sect:approaches}, along with an assessment of
how appropriate they are for problems of this paper. It turns out that more
accurate and informative methods can be made available. The method of renewal
strings \citep{storkey_renewal_strings} has been developed for overcoming the
difficulties of the standard approaches, and is ideally suited for detection
of satellite tracks, scratches and other linear phenomena. It is a Bayesian
probabilistic method, and so also provides some confidence measures for the
classification. Renewal strings are described in Sections~\ref{sect:renewal}
and~\ref{sect:fullmodel}. Their use is elaborated in
Section~\ref{sect:inference}. The results of applying these different methods
are shown in Sections~\ref{sect:detections} through to
\ref{sect:diffspikedetect} along with some analysis of performance. Evaluation
of the results is given in Section~\ref{sect:evaluation}. In conclusion,
information on how these results have been made available for the SSS data,
along with discussion of further work in this area and of the reciprocal
benefits of this work to the machine learning community can be found in
Section~\ref{sect:discussion}.

\section{Spurious objects in astronomical data}
\label{sect:spuriousdetail}A number of distinct classes of spurious object
commonly occur in optical/ near-infrared sky survey data. The descriptions
given here refer to the form they take within the SSS data. However many other
astronomical databases have similar characteristics.
\subsection{Satellite Tracks}
Satellite Tracks are due to movement of the satellite over the duration of
exposure for a given field. They follow paths which are almost straight
elliptic sections. Movement into or out of the Earth's shadow, the two ends of
exposure, or removal by the object recogniser can all stop the data related to
a satellite track from traversing the whole plate. The positions of satellite
tracks are unpredictable, and using a (probably incomplete) catalogue of
satellites and orbiting debris would be a complicated and probably unreliable
way of locating them. The data related to satellite tracks can vary
considerably. For some narrow and faint tracks the data can be sparsely
distributed along the track. For bolder tracks the data might consist of
objects with ellipses aligned along the track. Sometimes the data take the
form of very dense circular objects. Figure \ref{sattrackeg} gives two
examples of satellite tracks on the SuperCOSMOS Sky Survey plates and the
resulting data derived from them.
\begin{figure}
\begin{center}
\begin{tabular}{c}
{\bf Convert file md706\_fig1a.jpg to}\\
{\bf md706\_fig1a.eps for inclusion here}\\
 (a)\\
{\bf Convert file md706\_fig1b.jpg to}\\
{\bf md706\_fig1b.eps for inclusion here}\\
 (b)\\
\end{tabular}
\end{center}
{\caption{(a) and (b): Two tracks seen close up. Extracted data is shown as
ellipses superimposed on the digitised image. (a) A faint satellite track with
sparse spurious objects distributed along it. (b) A denser track with spurious
objects elongated along the track. The `blocky' appearance of the sky
pixels is a result of them having passed through a Haar--transform compression
algorithm. %
\label{sattrackeg}}}
\end{figure}

\subsection{Aeroplane Tracks}
Aeroplane tracks arise from aeroplane lights as they cross the field of view.
Often (but not always) the lights are flashing and so dashed tracks are seen.
All the representational issues which apply to satellite tracks also apply to
the data derived from aeroplane lights. Examples can be seen in
Figure~\ref{aertrackscratcheg}.
\begin{figure}
\begin{center}
\begin{tabular}{c}
{\bf Convert file md706\_fig2a.jpg to}\\
{\bf md706\_fig2a.eps for inclusion here}\\
 (a)\\
{\bf Convert file md706\_fig2b.jpg to}\\
{\bf md706\_fig2b.eps for inclusion here}\\
 (b)\\
\end{tabular}
\end{center}
{\caption{(a) A number of aeroplane tracks in field UKJ413:
the most vertical is a very solid aeroplane track, which has been converted
into very large elliptical objects. Some of the objects corresponding to this
track were
too large for pixel analysis and hence are effectively removed. The second
(sloping left) is a solid track converted into a large number of small objects
lying along the track. The third is one section of a dashed aeroplane track
corresponding to flashing lights. (b) A small slightly curved scratch.
Scratches are often longer, fainter or more curved than this, but small
scratches are also common. \label{aertrackscratcheg}}}
\end{figure}

\subsection{Scratches}
Scratches on the plate surface are not uncommon despite all the effort taken
to protect the emulsion from such. These scratches can be
seen by the SuperCOSMOS
digitiser as darker regions and hence are confused with photographic exposure.
They are usually (but not always) short, they tend to be curved, and sometimes
the curvature can vary significantly along the scratch. Again the same issues
occur in translating these linear features into data. An %
example of a scratch can be seen in Figure~\ref{aertrackscratcheg}.
\subsection{Dust fibres}
Fibres from clothing which contaminate the plate during scanning are not a
large problem in the original SSS data, but are a noticeable problem in the
SSS H--alpha survey, despite `clean room' %
operating conditions for the SuperCOSMOS measuring machine. This is because
the original photographic medium for the latter is film rather than glass, and
film is more prone to electrostatic attraction of particles. Although some of
the features of fibres can be removed by the methods developed here, many of
the very small tangled fibres of the H alpha survey would not be detected.
Although they are strictly one-dimensional features they tend to have many
discontinuities in their first derivative. They are often small, and the
combination of these effects mean they might only result in a few unaligned
objects in the derived dataset. Many of these fibres might be hard to locate
without going back to the original image data. See Figure~\ref{fibrehaloeg}
for an example.
\begin{figure}
\begin{center}
\begin{tabular}{c}
{\bf Convert file md706\_fig3a.jpg to}\\
{\bf md706\_fig3a.eps for inclusion here}\\
 (a)\\
{\bf Convert file md706\_fig3b.jpg to}\\
{\bf md706\_fig3b.eps for inclusion here}\\
 (b)\\
\end{tabular}
\end{center}
{\caption{(a) A small fibre on a plate resulting in 3 spurious objects. It
would be hard to detect these on database information alone. (b) A large
bright star on UKR005. the diffraction spikes and halos, along with their data
counterparts are clear. An outer halo is also evident, but in this case it is
too faint to have caused any detections. \label{fibrehaloeg}}}
\end{figure}
\subsection{Stellar Halos}
Because survey observations are optimised for faint objects, bright stars and
galaxies can often have annoying optical artefacts associated with them. The
halos around bright stars are the first of these which we will be considering.
These halos come from internal reflections within the telescope and take a
number of forms. First there is an area of brightness directly surrounding a
bright star, decaying with distance away from the star centre. Second there
could be a smaller uniform disc around the star which is more exposed than the
background. In the centre of a plate this disc will be centred at the star,
but at the edges it could be offset from the centre. This disc could have an
outer edge which is more exposed than the disc itself. Outside this inner disc
there might be another outer disc. This will be larger than the inner disc and
centred further from the star than the centre of the inner disc. Once again
this disc could have a more exposed outer edge. It is theoretically possible
to have further discs, but these are only occasionally observed. The discs are
elliptical.

When images containing these halo artefacts undergo pixel analysis,
there are generally two types of spurious record that are produced. %
First there is a high density of erroneous detections in the vicinity of the
bright star corresponding to the immediate bright area surrounding the star,
or in the region of the inner disc. Second there can be a ring of object
detections following the edge of either or both of the inner and outer discs.
Examples of this are seen in Figure~\ref{fibrehaloeg}.
\subsection{Diffraction Spikes}
Diffraction spikes are also associated with bright objects. They are almost
horizontal and almost vertical lines emanating from the bright star which are
due to diffraction about the telescope struts. The size and length of the
diffraction spikes is dependent on the brightness of the star: brighter stars
produce longer diffraction spikes. The deviation of the lines from the
horizontal and vertical is related to the position on the plate. Once again
greater deviations occur further from the field centre.

Because the SuperCOSMOS image analyser fits ellipses to objects, spurious
objects in the dataset along the line of the diffraction spikes often have
ellipses aligned along (or occasionally perpendicular to) the diffraction
spike. Examples are given in Figure~\ref{fibrehaloeg}.

\subsection{Other Detritus}
The vast majority of spurious objects fall into the above classes. However
other problems such as defects in the plate emulsion can produce spurious
objects in the dataset. Also small defects may not be detectable from only the
catalogue data alone and so it might be necessary to return to the original
images. This paper deals only with detecting spurious records from catalogue
data.

\subsection{Problems Caused by Spurious Objects}
Spurious objects will introduce errors in statistical results derived from the
data, and make locating particular classes of objects much harder. The fainter
tracks result in many spurious, elliptical low surface brightness `galaxies'
contaminating the respective object catalogue. A single--colour galaxy
catalogue, created from the UKJ survey for the purposes of studying faint blue
galaxies would therefore be highly contaminated by spurious, aligned image
records. This could severely impact a statistical analysis of the type
described in \citet{brown_galaxy_alignment}, where the degree and scale of
real galaxy alignment is being sought. In order to eliminate this possibility,
Brown et al. used a two--colour (JR) paired catalogue, but this of course
compromised %
the depth of the study and also biased it against faint blue galaxies which do
not appear on the R plates. Ideally, one might like to perform this study on a
single colour (J) galaxy catalogue. In many general problems we may be
interested in real objects which might be in one dataset but not in an other.
For example, objects which are evident at one wavelength but not at another
may be of interest. Fast moving stars will also be in different places in
catalogues derived from observations at different times, meaning that they
will not have exact positional matches across the datasets,
(e.g.~\cite{oppenheimer_position_matching}). Unfortunately satellite track
artefacts have the same characteristics, as they will only ever appear (in the
same place) in one dataset, and not in any other. Searches on non-matching
objects will bring up all the objects of interest \emph{plus} all of these
artefacts. When searching for rare objects the spurious records can be
overwhelming. Removing spurious objects, then, is of broad importance in
astronomy.

\section{Possible Approaches}
\label{sect:approaches} There have not been many attempts at tackling the
problem of labelling spurious objects derived from satellite tracks, scratches
or other linear phenomena despite the ubiquitous nature of the problem and the
difficulties these objects produce for many tasks which sky surveys are used
for.

\subsection{Hough Transform}
The most obvious way to locate lines of objects in two dimensional data
utilises the Hough transform. Indeed in \citet{cheselka_satellite_hough} and
\cite{vandame_satellite_hough} the authors followed this approach. The Hough
transform (\citealt{hough_transform}) is a standard image processing method
from which other related approaches have been developed. In its standard form
it is generally used in low dimensional situations to find lines containing a
high density of points hidden amongst a large number of other points
distributed widely across the whole space. Commonly it is used for line
detection in images.

The Hough transform works by moving from the space of points to the Hough
space, that is the space of lines. Every point $(d,\theta)$ in Hough space
corresponds to a line in the original space which is a perpendicular distance
$d$ from the centre of the data space and inclined at angle $\theta$ from the
vertical.

One method of implementing the Hough transform would search through a finite
number of line angles $\theta$. For each angle all the data points would be
considered. For each data point we would find the (perpendicular) distance
from the origin of the straight line \emph{through} that point at the relevant
angle. This distance would then be discretised, and the count in an
accumulator corresponding to this discretised distance would be increased by
one\footnote{A concise tutorial/demo of the Hough transform can be found at
http://www.storkey.org/hough.html}. The result of this is a count for each
angle and each perpendicular distance. Neglecting dependencies between the
accumulators at different angles and assuming, as a null hypothesis, a uniform
scattering of points\footnote{More formally assuming that points are sampled
from a homogeneous Poisson process.} in the data space, we know the
distribution of the count in a given Hough accumulator will be Poisson with a
mean proportional to the length of the corresponding line. If on the other
hand there is also a line of high density points in amongst the uniform
scattering, then this Poisson distribution will not be the correct model for
the Hough accumulator corresponding to this line. In fact the count will be
significantly higher than that expected under the null hypothesis. Hence
looking at the probability of the actual count under this null hypothesis, and
ideally comparing this to an alternative hypothesis based on some prior model
of line counts for satellite tracks, will indicate how likely it is that this
accumulator corresponds to a satellite track. A surprisingly large number of
papers on, and applications of, the Hough transform focus on finding large
absolute values contained within the Hough accumulators rather than comparing
them with the null distribution. Needless to say that approach is
significantly less accurate and powerful and is not to be recommended.

For an SSS dataset derived from field UKJ005, Figure~\ref{houghbin}a
illustrates the Hough transform of the data. In this figure lighter regions
correspond to higher accumulator counts. The large scale variation in light
and dark regions comes from the square shape of the plate: lines through the
centre along a diagonal are longer than off centre or off diagonal lines, and
hence will generally contain more stars. It is also possible to see some
sinusoidal lines of slightly increased intensity. These are caused by a local
cluster of large numbers of objects - either a galaxy or artefacts surrounding
a bright star.

The points in Figure~\ref{houghbin}a which have been circled correspond to
points which have an accumulator count much higher than that which would be
suggested by the null Poisson model. In fact one of these Hough accumulators
combined with its highest count nearest neighbours corresponds to the data
illustrated in figure \ref{houghbin}b. In this figure the data have been
rotated so that the horizontal axis shows the length along the line of the
Hough box (the region in data space corresponding to a given accumulator), and
the vertical axis corresponds to the much smaller combined width of the two
neighbouring Hough boxes. The representation shows that this Hough box does
indeed contain (part of) a satellite track, in fact the most prominent track
on the plate. Part of this track is illustrated in Figure~\ref{sattrackeg}(b).
\begin{figure}
\begin{center}
{\bf Convert file md706\_fig4a.jpg to}\\
{\bf md706\_fig4a.eps for inclusion here}\\
 (a)\\
{\bf Convert file md706\_fig4b.jpg to}\\
{\bf md706\_fig4b.eps for inclusion here}\\
 (b)\\
\end{center}
{\caption{(a) Hough transform of data from field UKJ005. The vertical axis
gives the distance from an origin in the centre of the plate (400 bins), the
horizontal axis gives the angle of orientation (200 bins). Lighter colours are
higher accumulator counts. The circled points are Hough accumulators with a
significantly high count, and which correspond to satellite tracks on the
plate. The original data which was accumulated in the 3 Hough accumulators at
points (0.38,[79 81]) of (a), that is points in the lower circle, is shown in
(b). Note the different scales of the two axes. The curvature of the track is
obvious from this plot. \label{houghbin}}}
\end{figure}

The curved shape of the track gives some hints of the problems which will be
encountered when working with satellite tracks. Many real stars and galaxies
lie within the the smallest Hough box which could contain the satellite track.
Hence flagging everything within the Hough box as possibly spurious will not
suffice. Reducing the size of the Hough boxes means that the data from a
single track will be split across a number of boxes, and the data within each
box might begin to be swamped by the general variations in underlying star and
galaxy distribution. This, combined with the fact that some of the tracks and
scratches we are interested in locating are very short segments means that the
data from the line can be swamped by the random variations in sample density
of all the other points along the line. Therefore nonlinear robust fits to the
data within Hough boxes are not enough. Add to this the problems of dashed
aeroplane tracks and the variable curvature of scratches and it becomes clear
that an approach is needed that is more flexible than the Hough transform.
Comparisons of the results obtained by the methods developed in this paper and
a Hough approach can be found in section \ref{sect:detections}.

\subsection{Elliptical Hough Transform}
\label{sect:ellipthough} Hough transforms can also be used for features other
than straight lines, although more than a few degrees of freedom increases the
Hough space which needs to be considered, and for large problems such as these
this would quickly become impractical. In fact even a standard circular Hough
transform, having three degrees of freedom to the Hough space (centre x coord,
centre y coord, radius) would be beyond reasonable computation for large
astronomical datasets. However if these degrees of freedom can be constrained
then the search space can be reduced to a reasonable size.

In the case of optical halos we know that the elliptical patterns are centred
at or near to bright stars, are axis aligned and are near circular. This
provides a significant constraint on the centre of the halo which is enough to
make an elliptical Hough transform entirely feasible for astronomical data.
The details of this particular implementation are given in
Section~\ref{sect:circhough}. In general, though, the elliptical Hough
transform is implemented in much the same way as the linear Hough transform.
First we decide on the Hough bin width, denoted $\epsilon$. The parameter set,
consisting of deviation from star centre, horizontal radius, and deviation of
vertical radius from a circle, are searched through. Each record in the
relevant locality is placed in the two accumulators corresponding to the
epsilon-width ellipses (with current parameters) which go through that point.
Again, after the process is completed, the expected count in each accumulator
is Poisson with mean proportional to the area of the corresponding ellipse. A
much higher count than that expected from this distribution would correspond
to an abnormally high density of points within that ellipse.

\subsection{RANSAC}
RANSAC (Random Sampling and Consensus, \citealt{fischler_ransac}) is a robust
estimation technique which is used when a large proportion of the data
provided is expected to be comprised of outliers. Unlike other robust
estimation techniques RANSAC does not use as much data as possible to obtain
an initial fit estimate. Rather it chooses a sample of as little data as
possible which will determine the required curve (2 points in the case of a
straight line). This sampling is repeated as many times as is necessary to
ensure that there is a high probability that one sample will obtain no
outliers. Each of these samples is then scored by calculating the number of
points that are no greater than a given distance $d$ away from the line. An
estimate of the line parameters can then be made from these points, or further
re-estimation methods can be used.

The RANSAC algorithm is simple. Suppose we are interested in fitting a
parametric curve/line with $k$ parameters, and there are $n$ data items.
Choose an acceptable probability of failure $P(\mbox{fail})$. Suppose we
expect there are $t$ items which will lie along the curve or line we want to
find and fit. The algorithm is
\begin{itemize}
 \item Repeat $s$ times
\begin{enumerate}
   \item Select $k$ data items.
   \item Fit the curve or line to these $k$ items.
   \item Calculate the support this curve has (i.e. how many other points lie along the
curve).
   \item Decide whether this curve is to be accepted or rejected dependent on
the support.
\end{enumerate}
 \item End repeat
\end{itemize}

Under these simple assumptions it is straightforward to show that we need
\begin{equation}
s=\frac{\log P(\mbox{fail})}{\log(1-t/n)}
\end{equation}
to get the required failure probability.

RANSAC is useful in situations where there are a large number of outliers.
However the situation presented here is one which exceeds the usefulness of a
naive RANSAC application. Given that we might be interested in finding lines
of $20$ data points in a dataset of $1$ million data points, that gives a
proportion of $1/50000$ of points not considered to be outliers. A naive
application of RANSAC would require a sample size $s$ of the order of
$(50000)^2$. Furthermore calculating the support for a curve would involve at
worst another run through the whole data, making the full cost $o(50000^3)$.

Less naively, a local RANSAC method could be developed. Most tracks are at
least piecewise continuous, and it is rare for large regions of tracks to be
unrepresented by an object in the catalogue. Recognising, then, that given a
true point (a point which is in the track) generally has another true point
within its $40$ nearest neighbours, say, reduces the required sampling size to
an order of $40 \times 50000$. However for each of these samples we would also
have to assess the quality of the local support for the points, which would
involve the further consideration of about $1000$ points in the local area to
assess whether they lie along the required line. Presuming we would be happy
with a $P(\mbox{fail})=1/100$ (where here this is the probability of detection
failure for each \emph{section} of track) this gives a cost of the order of $1
\times 10^{10}$ operations. A very accurate Hough transform considering $1000$
different angles will cost about $1\times 10^9$. Here we have neglected the
cost of finding the local neighbours. Again with this RANSAC approach,
accurately delineating the ends of a scratch requires further processing,
although the more local nature of the algorithm makes it easier. Local density
variations can also be included as this form of RANSAC involves local
assessment of support. The algorithm will be less accurate in situations where
large faint lines occur, as then the line will have to be recognised on the
basis of small amounts of local information alone, as there is no way of
accumulating information over larger distances. Also focusing on too local a
region can cause problems. Objects along a track will deviate from the track a
little, and if too %
short a distance is used to estimate the line of a track, the true track line
might never be found to enough accuracy.

In higher dimensions RANSAC becomes much more efficient. The Hough transform
scales exponentially with dimensionality, whereas dimensionality is irrelevant
for RANSAC. In general this makes RANSAC a more powerful technique.

The renewal strings algorithm of this paper is introduced and implemented in a
Hough-like framework. However it is a simple modification to implement it in a
RANSAC like framework. In this situation the line angle and positions are
chosen by sampling two points (ideally using the local form above), and
considering the line through the two points. The rest of the procedure remains
the same. With the data in this problem, we would expect the Hough-like and
the local RANSAC-like approaches to be the same order of magnitude in terms of
cost.

\subsection{Variations on the Hough transform}
The Hough transform has been part of the image processing toolbox for many
years, and it would be surprising if adaptations and advances had not been
made.
\subsubsection{Probabilistic Hough Transform.} That which has become known as
the probabilistic Hough Transform \citep{kiryati_hough_probabilistic} is
simply a way of using a subsample of the data to speed things up. It is
straightforward to calculate the probability of failing to detect a line that
would have been detected if all the data had been used. This can be used to
choose an appropriate number of points to subsample.
\subsubsection{Generalised Hough Transform.} If the feature to be detected is
not easily represented analytically, it might be possible to describe the
shape using a lookup table based on a prototype shape. The generalised Hough
transform \citep{ballard_hough_generalised} uses this approach.

\subsection{Related Work} In addition to the work already discussed,
there is a fair body of vision literature on robust techniques for line
segmentation. For example in \citet{kiryati_robust_lines} the authors use a
smoothing of the Hough accumulator (\citealt{hough_transform}) to obtain a
robust fit. However these approaches tend to be global straight line methods,
in the sense that they would not work well for either short line segments or
curved lines. In Cheng, Meer and Tyler
(\citealt{chen_robust_regression_mult_struct}), the authors provide methods
for dealing with multiple structures which need not cover the whole space.
Once again the model deals with straight line fits, %
and is tested on examples where there is not dominant background data, or
large numbers of outliers. Image based techniques for line extraction are
common, but tend to be based on continuity considerations, and they are not
appropriate in the context where we might be working with data derived from
images rather than the images themselves. The important work of Hastie and
Stuetzle (\citealt{hastie_principal_curves}) on principal curves provides a
different direction which does model curved data, but does not provide the
robustness and efficiency needed for situations when curves are set in large
amounts of other data.

\section{Renewal Strings}
\label{sect:renewal} Renewal strings, first introduced in the machine learning
literature in \citet{storkey_renewal_strings}, are a new probabilistic data
mining tool for finding subsets of records following unknown line segments in
data space which are hidden within large amounts of other data. The method was
developed specifically to address the problem of this paper. Renewal strings
combine a model for two dimensional data and a set of models for small numbers
of data lying on one dimensional manifolds within the two dimensional space.
The design of the model allows efficient line based techniques to be used for
separating out the data from the different one dimensional manifolds.

Renewal strings rely on a Bayesian generative approach, and so this section of
the paper starts by describing how renewal strings can be used to generate, or
simulate, artificial data of the sort we are interested in. Generative models
are a probabilistic framework, whereby a prior probability distribution is
built that represents belief about what data might be expected. This usually
involves forming a model where explanatory hidden (or latent) variables
provide a description of the data. The form of this model is such that
artificial data can be sampled from the prior distribution.

To use a generative model, it is inverted using Bayes' %
theorem to provide the
posterior distribution over the latent variables given the observed data. %
This posterior distribution can provide answers to any questions regarding
particular explanations for the data. Hence after the renewal strings model is
formulated, the same generative model is then inverted using the standard
Bayesian formalism to enable the key variables to be inferred from the
\emph{real} data. Although this inversion is approximate, it captures the
fundamental characteristics of the model.

Renewal strings depend on two tools of probabilistic modelling: the renewal
process and hidden Markov models. Hence these two models are introduced here.
\subsection{Renewal Process Description}
One way of modelling points along a line is through renewal processes. A
renewal process is a model for event times obtained by defining a probability
distribution for the time between events (commonly termed the inter-arrival
time). The time at which event $i$ occurs is dependent only on the time of the
previous event $i-1$; it obeys the Markov property. The typical example of a
renewal process is light bulb failure. The probability that a light bulb is
about to fail depends only on how long the light bulb has been burning (the
time since the last bulb failed) and not on the life (or any other
characteristics) of any of the other light bulbs which had been in the fitting
previously. Hence renewal models have the advantage that they are Markovian
while at the same time allowing complete flexibility in modelling the
inter-arrival times. As we will generally be dealing with one-dimensional
spatial concepts rather than temporal ones we will use the term `inter-point
distance' rather than `inter-arrival time' in the context of this paper.

It is possible that a Markovian model does not capture the major features of a
line process, for example a large inter-point distance might be much more
likely to be followed by a smaller inter-point distance (a problem
characteristic of bus arrival times, for example). The benefits of using a
Markovian model, in terms of speed and tractability, led to the decision to
focus exclusively on the inter-arrival characteristics of the data and ignore
any slight non-Markovian characteristics there might be. If this Markovian
model is not good enough then it can be possible to incorporate the
non-Markovian elements into the hidden Markov part of the renewal string
model, described in the next section.

\subsection{Hidden Markovian Dynamics} \label{sect:hmm} Hidden Markov Models
(HMMs) are a ubiquitous tool, seen in many different applications. Almost all
speech recognition systems use a hidden Markov model framework. They have also
been found to be a vital tool in gene sequence analysis, computer vision, time
series prediction and natural language processing. A standard introduction to
hidden Markov models can be found in \citet{rabiner_hmm}. In this section we
show how a hidden Markov model can be used to combine a number $m$ of
different sorts of satellite tracks or processes together into a switching
system.

Suppose we are given an inter-point distance $\Delta t_i$ at a point $i$ and
given prior renewal models for the $m$ different types of satellite track
processes. Then given a prior probability of a point being part of a
particular type of satellite track, we can obtain a posterior probability that
the inter-arrival time was characteristic of a particular type satellite
track:
\begin{equation}
P(X_i|\Delta t_i)=\frac{P(\Delta t_i|X_i)P(X_i)}{\sum_{X_i} P(\Delta
t_i|X_i)P(X_i)},
\end{equation}
where $X_i$ labels the type of process ($1,2,\ldots,m$ for the different types
of satellite track).

The problem with this is that the prior probability of a point being part of a
satellite track will be highly dependent on whether the last point in the line
was part of the same type of satellite track or not. Hence we need some prior
model for satellite track continuity. This is most easily defined using a
Markov model for the track labels $X_i$. Because $X_i$ are not observable the
whole model is called a hidden Markov model.

We introduce a set of conditional transition probabilities $P(X_i|X_{i-1})$
for the change in label between object $i-1$ and object $i$ along the line. We
also allow a transition $P(X_i=0|X_{i-1})$ where $X_i=0$ denotes the end of
the line.

The belief network for this system is illustrated in Figure~\ref{renewhmm}. In
a belief network, each node represents a random variable, and each directed
edge is a direct probabilistic dependence. Hence a belief network is an
implicit representation of the conditional independence structures in a
distribution. The nodes directly upstream from a given node are called the
parents of that node. For each node $V$ a conditional distribution
$P(V|\mbox{Parents of }V)$ needs to be defined. The joint probability
distribution over all the nodes is simply the product of the conditional
distributions for each node. For more details on belief networks see e.g.
\citet{castillo_expert_prob_network}. Probability models which can be
represented as belief networks without undirected cycles have the advantage of
allowing efficient exact inference to be done using belief propagation
\citep{pearl_firstbook}.
\begin{figure}
\begin{center}
{\bf Convert file md706\_fig5.jpg to}\\
{\bf md706\_fig5.eps for inclusion here}\\
\end{center}
{\caption{Belief network for the hidden Markov renewal process.
\label{renewhmm}}}
\end{figure}

The combination of renewal processes and hidden Markov models, henceforth
called renewal process hidden Markov models is not new within temporal
settings. It has been used for
(amongst other things) %
modelling the pecking behaviour of pigeons
\citep{otterpohl_pigeon_pecking_shortbib}! Also, in the case that the renewal
processes are all Poisson processes,  there is a direct relationship between
the Renewal Process hidden Markov model and the Markov modulated Poisson
process \citep{scott_markov_mod_poisson_proc}.

\subsection{Other Variables}
Though the positions of the objects will play the most important part in the
characterisation of line processes, other characteristics of the data might
well be able to contribute to the classification. For example, in satellite
tracks, the object ellipses tend to be aligned along the track.

We can modify the hidden Markov model to include input units to $X_i$ relating
to any other data fields associated with each point. This produces an obvious
generalisation to the hidden Markov model, known as an input-output hidden
Markov model \citep{bengio_iohmm}. Specifically we include elliptical
alignment information in the model.

\section{Renewal String Generation}
\label{sect:fullmodel} One way to visualise the complete renewal string model
involves building a background image of the stars and galaxies. Having decided
on the number and location of the satellite tracks, and the type of each, we
thread beads onto a string for each satellite track, where the distances
between the beads are defined by the hidden Markov renewal process, stopping
when we get $X_i=0$ in the hidden Markov model. Then we place the beads down
on to a background image, keeping the string tight. The final data consists of
the positions of the stars and galaxies in the background model, combined with
the positions of the beads.

More formally, the Renewal String generative model is built as follows. First
$2$ dimensional star and galaxy positions are generated from a background
spatial model. This could be any spatial process such as an inhomogeneous
Poisson process. For the purposes of this paper we define the background model
to be a Poisson process which is homogeneous within small regions, but has
different rates in different regions. Denote this rate function $\Lambda(\rB)$
for positions $\rB$.

Track processes are superimposed on the background data, to simulate satellite
tracks or scratches. There are potentially a number of different track
classes, each with different inter-point distributions. The tracks are
generated as follows:
\begin{itemize}
\item For each $\theta$ from a large but finite set of angles $\Theta$, and for
each of a finite set of lines $L$ at that angle, each of a given (narrow)
width $w$, a renewal process HMM is used to generate track data.
\end{itemize}
The renewal process hidden Markov model along the line is implemented using
this procedure:
\begin{itemize}
\item Along each line, a Poisson process (with large mean inter-point distance
$\gamma$) is used as a \emph{birth} process for the track; an event in this
process signifies the start of the track generation.
\item The class of the
track is chosen from the prior distribution $P(X_0)$,  and track points are
sampled by generating from a renewal process HMM: the inter-point distance is
sampled from $P(\Delta t_i|X_i)$ conditioned on the current class, and a new
point $t_{i+1}$ is placed the distance $\Delta t_i$ away from the current
point, at angle $\theta$ from that point. Then the next class is chosen from
the transitions $P(X_{t+1}|X_t)$.
\item We stop generating the track either
when the edge of the plate is reached or the hidden Markov chain transitions
into the `stop' class. The transition into the stop class initiates the birth
process again, which allows more than one track to be generated along the same
line.
\item Each point in each track is independently perturbed perpendicular to the
line of the track uniformly across the track width $w$.
\end{itemize}
Note that it is possible for a track to be turned on before reaching the
region of interest (in this case the plate edge), but not yet be turned off,
and hence the track will start at the edge of the plate. As the birth process
produces rare events most lines will not contain any tracks at all.

The set of angles $\Theta$ is generally chosen to be regularly spaced between
$0$ and 180 degrees, and the lines $L$ are chosen to cover the region of
consideration with a 2 line overlap; each point in the space lies in $2$ and
only $2$ lines at a given angle.

Figure \ref{generate} illustrates a sample from a generative model of this
form. We see a background model, along with two different types of tracks, one
of which is a high density broken line, the other a medium density line. The
model only generates straight line segments. Curves can be approximated using
piecewise linear segments.

\begin{figure}
\begin{center}
 \epsfxsize=8cm
{\bf Convert file md706\_fig6.jpg to}\\
{\bf md706\_fig6.eps for inclusion here}\\
 %\epsffile{md706_fig6}
\end{center}
{\caption{A sample from a 3 hidden-state renewal strings prior, illustrating a
background model, and 7 lines with differing characteristics.
\label{generate}}}
\end{figure}

\section{Inference and Learning}
 \label{sect:inference}
The generative Renewal String model says how to simulate background data and
the data corresponding to different tracks from a given parameter set. These
two elements can then be combined to form the final observable data. What we
are interested in is the inverse model: how to separate the background and
track records given the whole data and a particular set of parameters. This
inverse is given by Bayes theorem:
\begin{equation}
P(X|D,\Phi)=\frac{P(D|X,\Phi)P(X|\Phi)}{P(D|\Phi)}
\end{equation}
where $D$ denotes the data, $X$ a set with each element labelling whether a
point corresponds to a star/galaxy or one of the different types of track.
$\Phi$ is presumed to be a known set of parameters.

We require a reasonably fast inference scheme for this model. Borrowing from
the Hough transform approach it could be sensible to resort to line-based
techniques in order to perform inference. The Hough transform looks through a
comprehensive set of lines in the data, and finds those with a high
accumulator. To implement the renewal string we take this one step further.
Rather than just count the points along the line, a renewal process HMM is run
along the line to find points which could be best classified as part of a
track. This approach is only an approximate inference scheme for the
aforementioned generative model. The main issue is that, as with the Hough
transform, the dependence between lines at different angles is ignored. The
inference scheme for a single line is exact in the case where data from all
tracks other than those along that line have been removed. In reality, though,
such data remains. However because there are few tracks, tracks at other
angles will contribute at most a small number of points to the data along the
current line, and so this is likely to have limited effect on the inference
for the current line. This is the primary approximation assumption of the
inference method.

To work with lines rather than with spatial variables, we use the fact that a
spatial point distribution which is an inhomogeneous Poisson process will
correspond to an inhomogeneous Poisson process along the length of any line
(with some given width) going through that region of space. Hence when we
condition on the fact that we are considering one particular line, a one
dimensional Poisson process can be used instead of a spatial one. The
inhomogeneity of the Poisson process takes care of the fact that the
background model is not likely to take the same form across the whole plate.

Suppose we have an estimate for the density $D_b$ of background objects local
to each point. The full initialisation and inferential process can now be
given. As stated above, $\Theta$ gives the set of angles to be considered
(from $0$ to $180$ degrees), and $L=L(\Theta)$ the set of lines at each angle:
\begin{itemize}
\item[1)] Set the line width $w$ based on the expected maximum width of the lines to be
found. Define the inter-point distance distribution $P(\Delta t|X,D_b)$ for
each class $X$ including the background class. This can depend on the
background object density at that point. Define the class transition
probabilities $P(X^k_i|X^k_{i-1})$ and initial probabilities $P(X^k_0)$.
\item[2)] For each angle $\theta$ from the set $\Theta$
\begin{itemize}
\item[a)] For every point in the dataset, find all the lines $L'$ of width $w$ in $L$
which contain the point. Store the position $t$ \emph{along} each line in $l'$
in a bin corresponding to that line.
\item[b)] For each line in $L$, sort all the distances in its bin. Use these
distances as the data for an HMM with emission probabilities $P(\Delta
t|X,D_b)$ and transition probabilities $P(X^k_i|X^k_{i-1})$. Run the usual
forward-backward inference (see Appendix~\ref{append:forwardback}) to get
marginal posterior class probabilities for each point. Flag any points which
have a low probability of being background objects and note the angle at which
these points were detected.
\end{itemize}
\item[3)] End for
\end{itemize}
At the end of this process, the flagged points are the points suspected to be
part of a track or scratch. The associated probability gives extra information
regarding the certainty of this classification.

Note that, in terms of the generative model, the transition probability out of
the background state, $P(X^k_i \ne 0|X^k_{i-1}=0)$, is given by the
probability that the point is generated by the birth process rather than the
background process. In practice, at least for this work, we approximate this
by a fixed empirically determined value. Then we can take the initial class
probability $P(X^k_0)$ as given by the equilibrium distribution of the Markov
chain.

To estimate the rate $\Lambda(r)$ of the background inhomogeneous Poisson
process, we assume there is a length scale $s$ such that, for regions of size
$s \times s$, the contributions from the satellite tracks to the total number
of points, and the variation in background star/galaxy density are both
negligible. Then the local mean of the background Possion process can be
approximately obtained from the total density of points in a local region of
size $s \times s$.

Tuning of the parameters could be done with the usual expectation maximisation
algorithm for HMMs \citep{rabiner_hmm}. On the other hand empirical ground
truth estimates could be used to set the parameters. In this work the tracks
are also modelled as Poisson processes (a specific form of renewal process
with an exponential inter-point distance). The fundamental reason for this is
that along the line of a satellite track there will also be objects
corresponding to stars and galaxies. The point density along a track from a
satellite moving in front of a dense distribution of stars will be higher than
one passing in front of a relatively sparse region of sky, and hence the line
of objects along each track is a superposition. Poisson processes have the
advantage that the superposition of two Poisson processes is a Poisson
process. The equivalent statement is not true for more general forms of
renewal processes.

\subsection{Preprocessing using the Hough transform}
\label{houghpreprocess} Although we will show in section \ref{comphough} that
a standard Hough transform will not be able to find all the linear features in
the data, it is certainly true that it will find the larger satellite tracks
and other features that cover a significant distance on the plate. If these
larger features were the only ones of significant interest then the Hough
transform could be used as a preprocessor to determine which lines should be
checked for the features we want. The renewal string algorithm is only run
along a particular line if there seems to be enough support obtained from the
number of objects along that line. In this sense renewal strings are a
complement to the Hough transform rather than an alternative.

\section{Detections of satellite tracks, aeroplane tracks and scratches}
\label{sect:detections}A simple form of the renewal strings model was tested
on plate datasets within the SSS. For the background star/galaxy process the
local density was obtained by gridding the whole space into $40,000$ boxes and
counting the elements in each box. Improvements could be made through the use
of a k-means or other density estimate. $1000$ different angle settings were
used, and $18000$ different bins for the distance from the origin. Each data
point was put in two bins (i.e. the line width was twice the distance
separation). These values were obtained from simple geometric arguments. The
number of bins for the distance from the origin was chosen based on the
largest widths of the tracks which we were trying to detect. Then the angular
variation was then chosen such that any significant length of any track will
not be missed between two different angles.

A simple model of two hidden states was used, one corresponding to the
background, another to the satellite track. The inter-point distribution for
the satellite track was set to be an exponential distribution using the
empirical mean from a training set including $30$ different satellite tracks
from low density plates (the resulting mean was
$360$ microns on the plate, corresponding to 24~arcsec on the sky). %
As stars and galaxies also appear along satellite tracks, this empirical mean
was added to the mean of the background process to properly model the density
along a satellite track in different circumstances. The transition
probabilities were set approximately using prior knowledge about the number of
satellite tracks etc. on the training plates, the number of objects in total
and the number of objects per satellite track. This resulted in the transition
matrix $P(X_t|X_{t-1})$ for $X=\{background,\ track\}$ of
\[
\left(\begin{array}{cc}
0.999998&0.04 \\
2 \times 10^{-6}&0.96
 \end{array} \right)
\]
The initial prior probabilities were assumed to be the equilibrium
probabilities of a Markov chain with these transitions.

Figures \ref{vartracks} and \ref{vartracks2} gives a few examples of the
results. The whole plate is a little under $350$mm square, so some of these
images are for very small regions. Figure \ref{vartracks}b shows the results
for a whole plate. Note these images also contain the results for diffraction
spike and halo detection discussed in Sections \ref{sect:circhough} and
\ref{sect:diffspikedetect}. Stars or galaxies lying behind the path of a
satellite track are also flagged, as the characteristics recorded for those
objects will be affected by the existence of the track, and will therefore be
unreliable.

\begin{figure}
\begin{center}
\begin{tabular}{c}
{\bf Convert file md706\_fig7a.jpg to}\\
{\bf md706\_fig7a.eps for inclusion here}\\
 (a)\\
{\bf Convert file md706\_fig7b.jpg to}\\
{\bf md706\_fig7b.eps for inclusion here}\\
 (b)\\
\end{tabular}
\end{center}
{\caption{(a) An aeroplane track sloping from the bottom left to the top
right: a faint dashed line caused by a flashing light is properly detected.
(b) The detection results for the whole of UKJ159, including a number of
satellite tracks or aeroplane tracks. One flashing aeroplane track traverses
the right side of the plate. \label{vartracks}}}
\end{figure}

\begin{figure}
\begin{center}
\begin{tabular}{c}
{\bf Convert file md706\_fig8a.jpg to}\\
{\bf md706\_fig8a.eps for inclusion here}\\
 (a)\\
{\bf Convert file md706\_fig8b.jpg to}\\
{\bf md706\_fig8b.eps for inclusion here}\\
 (b)\\
\end{tabular}
\end{center}
{\caption{Detections along the parts of two satellite tracks on UKJ005. Other
parts of these tracks were illustrated in Figure \ref{sattrackeg}(a) the
sparse track and (b) the dense track. \label{vartracks2}}}
\end{figure}

\subsection{Comparison of results with naive Hough transform}
\label{comphough} The simple Hough transform does a slightly different job
from the renewal string approach, as it is designed to find lines which
traverse the whole plate. If we wish to find line segments we have to do some
post-processing of the results. The exact position of the tracks would still
need separating from the other points in that Hough box. Even so we can assess
how well the Hough transform can find lines which contain linear features.

There are many decision functions which can be used with the Hough transform.
For a useful comparison with the renewal string results, we look at the
significance level which would be needed to detect each track that was
detected with the renewal string method. We also look at how many other false
positive tracks would also be detected for given significance levels. The
number of angles and line widths considered were set to match the renewal
string settings ($1000$ angles, $9000$ different perpendicular distances)

Results for doing this on plate UKR002 are shown in Table~\ref{houghtest}.
This plate has no satellite tracks that traverse the whole plate, but does
have some smaller to medium size (a quarter of plate width) tracks and
scratches. Each track was located in a semi-automated way, and diffraction
spikes were ignored by removing all tracks within $1.5$ degrees of the
horizontal or vertical. The position and angle of each track was noted, and
included in a track list. In general each track was noted once. However, where
there was a large curvature to a track, more than one reference could have
been included in the list. The points corresponding to the plate notes in the
bottom left of the plate, and detections relating to a halo about a bright
star were removed by hand. This left $35$ tracks or scratches in the reference
list.

For comparison purposes, we looked at all the listed tracks and calculated
what significance level would be needed in order to detect the line containing
that track with the Hough transform. The table shows the significance level
required to detect the tracks along with the total number of tracks (true and
false) which would have been detected by the Hough transform at various
significance levels. These counts once again exclude Hough accumulators
corresponding to lines within $1.5$ degrees of the horizontal or vertical. The
result was that a total of $968$ different angles were considered.
Accumulators with an expected count less than $12$ were discarded as these are
easily affected by isolated points.

Many of the tracks are picked up by the Hough transform for high significance
levels. However some of the tracks are not even detectable at significance
levels of $0.5$ and smaller. Hence the renewal string approach is certainly
increasing the detection rate compared with using the Hough transform alone.
Furthermore the Hough transform produces large numbers of false positives even
when only choosing very significant lines. The number of false positives on
this plate is much greater than the theoretical number that should be found at
the high significance levels. Some of these will be contributions from
accumulators mapping to lines overlapping a track at a slight angle. However a
dominant reason for the discrepancy is that global approaches like the Hough
transform do not easily deal with variations in the background density; there
is an assumption of homogeneity. If many stars are clustered in one location,
then they can cause a significant contribution to a single Hough accumulator.
\begin{comment}
An alternative approach involves searching for peaked local maxima in Hough
space. This can reduce some of the problems as local star clusters cause false
positive detections in many adjoining hough boxes. However there is no
principled way of making decisions regarding significance within this
heuristic framework, and even then a large number of false positives would be
detected.
\end{comment}
As mentioned in section \ref{houghpreprocess}, if only the more significant
detections are wanted then the Hough transform can be used to find proposal
search lines, and then the renewal string approach allows the exact points in
the track to be found along that line (if there are any). This can be a
significant speed up over running a hidden Markov renewal process along every
line. How many tracks would be missed depends on the significance level used,
and in this circumstance can be estimated from Table~\ref{houghtest}. The
lower the significance level, the more lines that would have to be checked,
and hence the greater the computational cost.

\begin{table}
\begin{center}
\begin{tabular}{|r|r|r|r|}
 SIGLEV&DET&TOT&THEOR\\ \hline
 $0.5$&$31$&$3.18\times 10^6$&$4.4\times 10^6$\\
 $0.7$&$26$&$2.09\times 10^6$&$2.6\times 10^6$\\
 $0.9$&$21$&$8.96\times 10^5$&$8.7\times 10^5$\\
 $0.95$&$15$&$5.34\times 10^5$&$4.4\times 10^5$\\
 $0.99$&$9$&$1.62\times 10^5$&$8.7\times 10^4$\\
 $0.999$&$7$&$30147$&$8712$\\
 $1-10^{-4}$&$5$&$5903$&$871.2$\\
 $1-10^{-5}$&$4$&$1158$&$87.12$ \\ %
 $1-10^{-6}$&$3$&$257$&$8.7$\\
 $1-10^{-7}$&$2$&$71$&$0.87$\\
 $1-10^{-8}$&$2$&$25$&$0.087$\\
 $1-10^{-9}$&$1$&$16$&$0.0087$

 \end{tabular}
\end{center}
{\caption{The number of the $35$ tracks/scratches on UKR002 which would have
been detected using the hough transform. SIGLEV gives the significance level
used. DET gives the number of the tracks which would have been flagged at that
significance level, TOT the total number of lines flagged as significant by
the Hough transform, and THEOR the theoretical number of false positives for a
homogeneous Poisson distribution. A significance level of $1-10^{-7}$ is
needed to reduce the theoretical false positive detection rate to a suitably
low level. Then only two of the tracks could have been detected, and in
practice there would have been many false positives flagged.
\label{houghtest}}}
\end{table}

\section{Detections of optical halos}
\label{sect:circhough} Finding optical halos is possible using elliptical
Hough transforms described in section \ref{sect:ellipthough}. As the halos are
almost circular, and centred near to bright stars it is only necessary to
consider ellipses up to a certain radius, and in a limited number of centres,
and with a limited amount of ellipticity. The possible centres are chosen to
be near to bright stars.

To search around bright stars, a bright star set is needed. The measurement of
the photographic magnitude of bright stars can be subject to quite large
error. For this reason the measured size of the star is used as an indicator
of its likelihood of producing halos (or diffraction spikes). We chose to
consider all stars with a measured radius greater than $200$ microns. The
star/galaxy classification flag can also be inaccurate for very bright stars,
and so any object which is approximately circular is presumed stellar. We
allowed a minimum ratio of $0.7$ between the minor and major elliptical axes.
This does result in some misclassification where stars in circular galaxies
are presumed to be part of a halo around a star. Solutions to this problem are
being investigated, including the possibility of building more accurate
classifiers for stars and galaxies by training on the Sloan Digital Sky Survey
classifications.

To detect the halos an elliptical Hough transform was used. The elliptical
axes were presumed to be aligned along the x and y axes, and the ratio of the
horizontal and vertical axes was varied from $0.8$ to $1.2$ in intervals of
$0.05$.
The ellipse centres were chosen to be within $400$ microns of the measured
star centre, stepping in $80$ micron intervals. The Hough transform searched
through $200$ different radii, each of width $40$ micron. An allowance for the
variation in background density of $1.1$ times the measured density was used.
The halos were expected to have a mean line density of $1$ point every $380$
microns over and above the background density, obtained from observations
regarding the density of halos. Empirical estimates were used to calculate a
prior probability of a randomly chosen ring about a bright star containing a
halo of about $0.00003$.

The objects were flagged when any ellipse containing them was found to have a
 posterior probability of greater than $0.5$ of being a halo. The posterior probability
of an object being part of a halo was assumed to be the greatest posterior
probability of all the ellipses containing that object.

\section{Detection of diffraction spikes} \label{sect:diffspikedetect}
Diffraction spikes occur around bright stars. They are linear features with
many of the characteristics that other linear features have, and it is true
that the application of the standard renewal string approach of the previous
section will detect a large number of the diffraction spikes without any
modification. This is because the renewal string approach is particularly
suited to detection of short linear features as it is based on a model which
allows the generation of short lines.

Despite the fact that diffraction spikes are also linear phenomena similar to
scratches, there are significantly fewer degrees of freedom regarding where
they lie. Hence they can be found with greater accuracy by focusing
exclusively on lines passing through bright stars, and aligned almost
horizontally and almost vertically to the image axes. This means that the
renewal string methods described in the previous sections can be enhanced by
restricting the renewal string algorithm to look only at near vertical and
near horizontal lines in the region of a bright star. This enables the
probability model to be tailored specifically to diffraction spikes rather
than to all linear features.

The restricted renewal string approach was used in order to try to pinpoint
the diffraction spike positions more accurately. The bright star set described
in section \ref{sect:circhough} was used. $17$ different angles in $0.3$
degree gradations, and $17$ different line positions at gradations of $13$
micron were considered near to the axis aligned lines through the centre of
the bright stars. These values appeared to cover the variation in the position
and angle of the spikes without introducing excessive computational burden. In
a similar way to the renewal string model, a hidden Markov renewal process was
run along each of these lines. The main difference is that the process started
at the closest point to the star centre, working out to the edge. The mean of
the spike Poisson process was taken to be 190 microns, and the transition
probabilities $P(X_t|X_{t-1})$ for $X=\{background,\ track\}$ were
\[
\left(\begin{array}{cc}
0.9992&0.23 \\
0.0008&0.77
 \end{array} \right)
\]
Due to the increased probability of a point near to the star being a part of a
spike, the initial probabilities are no longer the equilibrium probabilities
of the Markov chain. The initial probabilities $P(X_0)$ were set to be
\[
\left(\begin{array}{c}
0.994 \\
0.006
 \end{array} \right)
\]
These probabilities were estimates based on the number of lines considered,
the expected number of diffraction spikes which existed per star examined, the
overlap of the lines, and the expected length of the lines. It is possible
that these hand estimates could be enhanced using the expectation maximisation
(EM) algorithm to obtain maximum likelihood parameter estimates. However that
would increase the computational burden significantly for what would probably
be small gain.

The usual renewal string inference (the forward backward equations of appendix
\ref{append:forwardback}) was used to detect the positions of the diffraction
spikes, again flagging for posterior probabilities greater than $0.5$.

\section{Evaluation}
\label{sect:evaluation}The detections were evaluated by an astronomer (NCH),
who looked through a printed version of the plate data for a whole plate
(UKR001). The plate was split into $36$ regions, each region being printed on
an A3 sheet. These A3 sheets were examined closely for false negative and
false positive detections, and the astronomer also commented on other aspects
of the detection he felt notable. In this analysis features corresponding to
small fibres were ignored. As the measured characteristics of true stars or
galaxies along or very near a satellite track will be affected by the track,
these objects should also be flagged.

A general summary of the results can be found in Table~\ref{fpfnres}. All the
major satellite tracks were found and the ends of the tracks were generally
accurately delineated. All of the small scratches were properly identified,
although one of them involved a significant bend. Figure
\ref{curvedscratchUKR001} illustrates this. Some of the objects along the bend
were improperly classified as real objects. A small number of small false
positive linear detections were made. Some objects due to fibres on the plate
were also picked up, although as expected the method was not designed for and
is not ideally suited to their detection.

\begin{table}
\vspace{1cm}
\begin{center}
\begin{tabular}{|l||r|r|r|r|r|r|}
 &FP&FP\%&FN&FN\% &DET&TOT\\ \hline \\
 Tracks&60&0.7&14&0.0033&8539&429238 \\ \\
 Halos&60&1.2&32&0.0075&5063&429238\\ \\
 Spikes&30&2.0&175&0.04&1482&429238
 \end{tabular}
\end{center}
\vspace{1cm}

{\caption{Numbers of false positive (FP) and false negative (FN) records for
satellite track/scratch detection, halo detection and diffraction spike
detection on plate UKR001. False positive percentage expressed as percentage
of total detections (DET); false negative percentage expressed as percentage
of [total objects (TOT) - total detections (DET)]. \label{fpfnres}}}
\end{table}

\begin{figure}
\begin{center}
\begin{tabular}{c}
{\bf Convert file md706\_fig9a.jpg to}\\
{\bf md706\_fig9a.eps for inclusion here}\\
 (a)\\
{\bf Convert file md706\_fig9b.jpg to}\\
{\bf md706\_fig9b.eps for inclusion here}\\
 (b)\\
{\bf Convert file md706\_fig9c.jpg to}\\
{\bf md706\_fig9c.eps for inclusion here}\\
 (c)
 \end{tabular}
\end{center}
{\caption{(a) False positives (FP) and false negatives (FN) for detections
along a very faint highly curved scratch on plate UKR001. This is the only
significant source of false negatives for scratch/track detection that the
astronomers found on this plate (the others were isolated points at the end of
a scratch). The scratch is not easily seen in the corresponding image (b), but
can be seen in the detail (c) of the brightened region of (b).
\label{curvedscratchUKR001}}}
\end{figure}

Another useful evaluation involves comparing the detections with cross-plate
matches. Any objects which are due entirely to a satellite track or scratch
will not have a corresponding record in other data taken at different epochs
or different wavelengths. If we presume that any objects which do pair across
different surveys are real astronomical objects or optical artefacts, then we
can make an assessment of how many objects are being flagged which are
definitely not due to a track or scratch.

\begin{comment}
The approach was tested on eight test plates and the results compared with
those found by examination of the plate data by eye. Furthermore the results
were compared with match data which paired stars and galaxies from surveys at
different wavelengths. Objects which are the result of satellite tracks will
be unmatched. In this analysis we ignore any diffraction spikes found by the
renewal string algorithm.

All $38$ tracks and scratches found by eye were also found by the algorithm.
Many of these were small curved tracks. On three of these tracks there was an
overshoot error, where the renewal string approach predicted a few objects
beyond the actual end of the satellite track. The renewal string also
incorrectly flagged $11$ other tracks as satellite tracks. These were all
short tracks containing less than $20$ objects. A further $3$ turned out to be
short scratches not detected by eye.
\end{comment}
Of the $3.1$ million objects on the $8$ plates, $10029$ objects were located
by the renewal string algorithm as part of a scratch or track. Of these only
$552$ ($5\%$) were paired across different surveys, indicating they were true
astronomical objects or related to optical artefacts. Nearly all of these were
stars and galaxies lying along the line of the satellite tracks, which should
be flagged as problematic anyway.

Some examples of the tracks found can be seen in Figure~\ref{varioustrackeg}.
Figure \ref{notfalsepos}a shows a number of points which were flagged as
spurious by the renewal string approach. This is located on UKR001 around
$RA=2:34:52$, $DEC=-86:32:16$. The astronomer marked this up as a false
positive detection, as it did seem to look like stars and galaxies which just
happened to be aligned. However when looking at the image (Figure
\ref{notfalsepos}b), it is clear that these points are aligned along a very
faint track. For most of the track the image is too faint to produce any
spurious records. However for this short section some spurious records do
occur. The fact that this is the case can be seen by looking at the
corresponding image part in the overlap with plate UKJ003, where no such
objects are recorded.

\begin{figure}
\begin{center}
\begin{tabular}{c}
{\bf Convert file md706\_fig10a.jpg to}\\
{\bf md706\_fig10a.eps for inclusion here}\\
 (a)\\
{\bf Convert file md706\_fig10b.jpg to}\\
{\bf md706\_fig10b.eps for inclusion here}\\
 (b)
 \end{tabular}
\end{center}
{\caption{(a) A very short scratch which was detected. (b) A typical part of a
large satellite track. \label{varioustrackeg}}}
\end{figure}

\begin{figure}
\begin{center}
\begin{tabular}{c}
{\bf Convert file md706\_fig11a.jpg to}\\
{\bf md706\_fig11a.eps for inclusion here}\\
 (a)\\
{\bf Convert file md706\_fig11b.jpg to}\\
{\bf md706\_fig11b.eps for inclusion here}\\
 (b)
 \end{tabular}
\end{center}
{\caption{(a) A set of detections on UKR001 which were marked up as false
positives by the astronomers. (b) A look at the image shows the points are in
fact part of a faint track. \label{notfalsepos}}}
\end{figure}

Again on UKR001, the halos of the bright stars were picked up accurately.
There were some false positives due to the other local features being
misinterpreted as halos. For example a high density cluster might contribute
to a high Hough accumulator count for a given ellipse, causing other points in
that ellipse to also be classified inaccurately. In general though because the
halo detection is a Hough approach, the density variation along the path of
the larger halos is not taken into account, and this can cause problems such
as these. From Table \ref{fpfnres}, it can be seen that the number of false
positives is still a small proportion of the total detections
\footnote{Classifications are only illustrated for the deblended objects. Some
of the objects on these plots can be seen to be larger parent objects which
have a number of deblended children.}.

Most, but not all of the diffraction spikes were detected. In general we
presume that a diffraction spike will need to contain about $4$ objects before
we would expect this algorithm to detect it. The most common failures were
false negatives. The majority of these were diffraction spikes on stars which
were only just bright enough to have spikes, where the spikes were represented
as four objects. Some were failures to recognise the true extent of a
diffraction spike. Even so the number of false negatives was relatively small.
There were also some false positives due to overrun of the diffraction spike
detection beyond the end of the spike, and other effects such as two adjacent
bright stars whose halos and diffraction spikes interacted with one another.
Again the number of false positives was a small percentage of the overall
detections.

Examples of the halo and diffraction spike detection can be found in
Figure~\ref{halodiff}. These show one bright (a) and one less bright (b) star,
and their associated halos and diffraction spikes. Example detections on a
whole plate can be seen in Figure~\ref{vartracks}b.
\begin{figure}
\begin{center}
\begin{tabular}{c}
{\bf Convert file md706\_fig12a.jpg to}\\
{\bf md706\_fig12a.eps for inclusion here}\\
 (a)\\
{\bf Convert file md706\_fig12b.jpg to}\\
{\bf md706\_fig12b.eps for inclusion here}\\
 (b)\\
\end{tabular}
\end{center}
{\caption{(a) A set of halos and diffraction spikes around a bright star on
UKR002 at RA$=23:52:7$, DEC$=-82:01:14$. (b) One of many medium-bright stars
on UKR002 with associated halo and diffraction spikes. \label{halodiff}}}
\end{figure}

In general we are getting good detection rates for all three problem features.
This will make a major difference to the reduction in spurious data problems
in the catalogue.
\begin{comment}
\section{Speed and memory}
With data of this size, the speed and memory use of any algorithm is an issue.
Running the algorithm as it stands on an average plate takes about $??$ minute
of CPU time and uses $??$MB of memory. The provision of the Blue Dwarf pseries
machine from IBM has enabled runs on the whole SSS data to be achieved in a
few days.
\end{comment}
\section{Discussion}
\vspace{-1mm} \label{sect:discussion}
Renewal Strings have certainly aided the process of detection of spurious
objects in astronomical data: given very large amounts of data only a small
number of detections were made, most of which were correct. The form of the
model allows the use of the hidden Markov models and renewal processes,
resulting in a model that is efficient even for huge datasets. It has been run
all the plates of the SSS data (over $3000$ in total), providing a valuable
resource to astronomers.

Renewal strings are a practical, probabilistic approach to a large problem
requiring high accuracy. Renewal strings go beyond a local Hough transform
method to a general approach for detecting line segments within large amounts
of other data. Slightly curved lines are also detectable as a set of locally
linear parts.

Hough transform approaches were also suitable for halo detection. Some of the
false positives reflected the difficulty that Hough approaches have in dealing
with local density variations. One way of improving the current approach would
use renewal strings around the arc of the halo in much the same way that is
currently used for straight lines. This would allow the local densities to be
modelled more accurately.

The renewal string approach has also been adapted for diffraction spike
detection, and shows promising results. One noticeable improvement would
involve the introduction of prior information regarding the length of the
diffraction spike depending on the brightness of the star. However
inaccuracies in the measured star brightness, and significant variations in
spike length depending on the position of the star, or density and colour of
the plate, have made this a nontrivial task. Despite this the current method
is providing accurate detection results, and enables the recognition of the
vast majority of diffraction spike objects with relatively few false
positives.

The renewal string approach shows clear benefits over Hough approaches, and
has proven a highly effective method for detection of spurious data in the
SuperCOSMOS Sky Survey. The results of the method will reduce the problem of
spurious data in these surveys to insignificant levels. Furthermore the
technique is general and can be adapted for use in future sky surveys. The
techniques will also be useful in fully digital sky surveys. These techniques
are particularly suitable for detection of the shorter satellite and aeroplane
tracks which can be found in many digital surveys.

The results of the application of this approach to the SuperCOSMOS Sky Survey
will be made available in a forthcoming new release (Hambly et al., 2003, in
preparation) of the survey data. This new release will incorporate several
new data enhancements (eg.\ to proper motions, photometric calibration
scales and source pairing) along with enhancements to user access. The
existing standard SSS distribution is available at
\verb+http://www-wfau.roe.ac.uk/sss/+.

\subsubsection*{Acknowledgements}
\vspace{-1mm} This work is part of a project funded by the University of
Edinburgh. The authors also thank IBM for the generous provision of the
P-Series machine Blue Dwarf to the School of Informatics, Edinburgh through
the Shared University Research Programme. This machine was used for some of
the runs on the SuperCOSMOS Sky Survey data. The first author would like to
thank Microsoft Research for fellowship funding for the final stages of this
work.

\appendix
\section{Inference in Hidden Markov Models}
\label{append:forwardback}In this Appendix the update equations for a hidden
Markov model are given. For more details see e.g. \citet{rabiner_hmm,
castillo_expert_prob_network}. Suppose the number of states for the HMM is
denoted by $M$, the latent class variables denoted by $X$ and the visible
variables by $Y$. Subscripts are used to denote the time index from $0$ to
$T$. Then the state transition matrix at time $i$ is $P(X_{i+1}|X_i)$. The
output distribution is $P(Y_i|X_i)$, and the initial class probabilities are
$P(X_0)$. The update equations for inference in a hidden Markov model consists
of a backward and a forward pass. We presume that all the $Y_i$ are given. The
backward pass propagates the data likelihood back through time. Once that is
complete, the forward pass propagates the prior information forward through
time.

Let $\lambda_i(X_i)=P(Y_{i-}|X_{i})$ denote the backward message at time $i$
for each class ($1$ to $M$) taken by $X_i$. Here $Y_{i-}$ denotes the set of
all of the observable values for times after and including time $i$. Likewise
$\rho_i(X_i) \propto P(Y_{i+},X_i)$ is the forward message at time $i$, where
$Y_{i+}$ denotes the set of all of the observable values for times before time
$i$. Then we can update $\lambda$ by using the initialisation
$\lambda_T(X_T)=1 \ \forall X_T$ and then applying the recursive formula
\begin{equation}
\lambda_i=P(Y_i|X_i) \sum_{X_{i+1}} P(X_{i+1}|X_{i})\lambda(X_{i+1}).
\end{equation}
Likewise we initialise $\rho_i(X_0)=P(X_0)$ and apply the recursive formula
\begin{equation}
\rho_i(X_i)=\sum_{X_{i-1}} P(X_{i}|X_{i-1}) \rho_i(X_{i-1})P(Y_{i-1}|X_{i-1})
\end{equation}

The final marginal posterior probabilities (beliefs) are given by
\begin{equation}
P(X_i|\{Y_j \ \forall j\})=\alpha \lambda_i(X_i) \rho_i(X_i)
\end{equation}
with $\alpha$ a normalisation constant.

The beliefs can be calculated in time linear in the number of nodes.
\bibliography{fullbib,astrobibb}

\end{document}